\documentclass[twocolumn,showpacs,preprintnumbers,amsmath,amssymb]{revtex4}

\usepackage{bm}
\usepackage{graphicx,amssymb,amsmath,amsfonts}

\newcommand{\tr}{{\rm tr}} 

\newcommand{\ve}[1]{\bm{#1}} 

\newcommand{\Fne}{F^{\boldsymbol{n}, \epsilon}}
\newcommand{\Fme}{F^{\boldsymbol{m}, \epsilon}}

\newcommand{\wne}{w_{\ve{n}, \epsilon}}
	
\newcommand{\C}[1]{C(\ve{#1},\epsilon)}	
\newcommand{\id}{\openone}

\newtheorem{definition}{Definition}
\newtheorem{proposition}{Proposition}
\newtheorem{theorem}{Theorem}

\begin{document}

\title{ 
\begin{flushright}{\footnotesize  \textmd{{\em Phys. Rev. Lett.}} {\bf 88}\textmd{(24), 240402, 2002}} \end{flushright}

\vspace{.4in}
The Kochen-Specker Theorem for Finite Precision Spin 1-Measurements}

\author{Thomas Breuer}
\email{Thomas.Breuer@fh-vorarlberg.ac.at}
 
\affiliation{Department of Computer Science, FH Vorarlberg, Achstrasse 1, 
A-6850 Dornbirn, Austria}

\date{\today}

\begin{abstract}
Unsharp spin 1 observables arise from the fact that a residual 
uncertainty about the actual orientation of the measurement device remains. 
If the uncertainty is below a certain level, and if the distribution of measurement errors is covariant under rotations, a Kochen-Specker theorem for the unsharp spin observables follows: There are finite sets of directions 
such that not all the unsharp spin observables in these directions can consistently 
be assigned approximate truth-values in a non-contextual way.  
\end{abstract}
\maketitle

The Kochen-Specker (KS) theorem establishes that not all measurement outcomes predicted
by quantum mechanics can result from detecting hypothetically predetermined values of the
observables. Pitovsky \cite{Pitovsky 1985}, Meyer \cite{Meyer 1999}, Kent \cite{Kent 
1999}, and Clifton and Kent \cite{Clifton and Kent 2000} (MKC) claimed that for finite precision measurements the KS-theorem is irrelevant.  

Infinite precision is crucial to the KS-argument because non-contextuality 
can be exploited only if two measurements intended to pick out the same observable as member of
two different maximal sets pick out exactly the same observable. MKC show that 
non-contextual hidden variable models can be constructed if we relax the assumption of
infinite precision by an arbitrarily small 
anmount. In these models it is not exactly the observables in a KS-set that are 
assigned non-contextual values, but observables arbitrarily close to them.
In fact it is possible to assign values to a dense set of observables, namely to the spin
observables in directions with rational coordinates. 
So the hidden variable-theorist is free to adopt the hypothesis that due to some 
apparatus misalignment instead of the intended observable he measures  
another observable, which cannot be distinguished from the intended observable
by a finite precision measurement. MKC show that the results of this kind of
finite precision measurements can be explained by a non-contextual hidden variable model. 
While the mathematical results of MKC are correct, their nullification interpretation was 
widely questioned \cite{Cabello et al 1997, Cabello 2001, 
Simon et al 2000, Simon et al 2001, Mermin 1999, Havlicek et al 2000, Appleby 2001}.  

In this Letter we establish a KS-theorem for finite precision spin 1 measurements with
covariant error distributions. This result highlights the price to pay for MKC-type hidden
variable theories: in these models the error distributions are not covariant. Therefore
the spin observables describing the measurement statistics of these models are not angular
momentum observables. They do not obey the right commutation relations and their squares
do not have dispersion-free values. (Thus in the sense of the spin observables describing the measurement statistics of the MKC-models one could not speak of a ``spin 1"  particle.) 

Quantum computation promises algorithms which require only poly(log) number 
of bits precision for problems known not have polynomial time classical solutions 
\cite{Shor 1997}. Meyer \cite{Meyer 1999} claims his result implies that the KS-theorem 
alone cannot support this promise of quantum computation because the KS-theorem 
requires infinite precision for exhibiting the difference 
between quantum and classical systems. The present KS-theorem 
for finite precision measurements can account for the 
difference between finite precision quantum and classical computing.      

Summing up: KS exclude non-contextual hidden variable models of infinite precision quantum
measurements. MKC provide non-contextual hidden variable models of finite precision
quantum measurements. The present result excludes hidden variable models of finite
precision quantum measurements with covariant error distributions.  

\section{\label{section-Finite Precision Measurements of Spin 1 Observables}
Finite Precision Measurements of Spin~1 Observables}
An experimenter who wants to 
measure spin in a direction $\ve{n}$ will have a procedure for trying to 
do this as exactly as possible. Simon {\em et al.} \cite{Simon et al 2001} refer to this 
procedure by saying he sets the ``control switch" of his apparatus to the position 
$\ve{n}$. The switch position is all the observer knows about. In an operational sense, 
the physical observable measured is entirely determined by the switch position. However,
there will usually be some degrees of freedom of the apparatus which the experimenter cannot
control. This results in an apparatus misalignment of which the experimenter is not aware. 
If he were aware of it he would correct it. Not being aware of the 
misalignment he interprets the outcome produced by the misaligned apparatus as result
of an experiment without misalignment. 
Unlike in Simon {\em et al} \cite{Simon et al 2001}, in this paper 
the misalignment will not be described by associating hidden variables to the apparatus. 
Instead, the effects of the misalignment are described by unsharp spin observables. 

The sharp spin 1 observables in $\ve{x}, \ve{y}$, and $\ve{z}$  
direction are given by the three 
three-dimensional Pauli matrices $S_{\ve{x}}, S_{\ve{y}}, S_{\ve{z}}$, 
each of which has eigenvalues 1, 0, and -1. 
For example $S_{\ve{z}}$ is given by 
\begin{equation}
S_{\ve{z}} = 
\left(
\begin{array}{ccc}
1 & 0 & 0 \\
0 & 0 & 0 \\
0 & 0 & -1
\end{array}
\right). 
\end{equation}
Denote the eigenvectors of the spin matrix $S_{\ve{z}}$ by 
$\psi_{\ve{z},1},\psi_{\ve{z},0},\psi_{\ve{z},-1}$ and 
the  corresponding eigenprojectors by   
$P_{\ve{z},i}:=\vert\psi_{\ve{z},i}\rangle\langle\psi_{\ve{z},i}\vert$. 
The $P_{\ve{z},i}$ are sharp spin properties. For example, 
\begin{equation}
\label{eq-Pz1}
P_{\ve{z},1} = 
\left(
\begin{array}{ccc}
1 & 0 & 0 \\
0 & 0 & 0 \\
0 & 0 & 0
\end{array}
\right). 
\end{equation}
Similar notation will be used for the $\ve{x}$- and the $\ve{y}$-axes. 

For an arbitrary direction $\ve{n}$ the sharp spin 1 observable is 
$S_{\ve{n}}:=\ve{n}\cdot \ve{S}$, where $\ve{S}$ is the Pauli vector $(S_{\ve{x}}, 
S_{\ve{y}}, S_{\ve{z}})$.
$S_{\ve{n}}$ also has eigenvalues 1, 0, and -1. Let $\psi_{\ve{n},i}$ and $P_{\ve{n} ,i}$ be 
the eigenstates and eigenprojectors of $S_{\ve{n}}$ corresponding to the eigenvalues 
$i=1,0,-1$.  
The sharp spin observable $S_{\ve{n}}$ in direction $\ve{n}$ can be represented as a 
projection valued (PV)-measure on the value space $\{1,0,-1\}$, which associates
to each element $i$ of the value space $\Omega$ the projector $P_{\ve{n}, i}$.

Now assume we are not sure that we actually measure the spin in the intended direction 
$\ve{n}$. 
We only know that the directions $\ve{m}$ of actual spin measurements are distributed with 
a density $\wne(\ve{m})$ around $\ve{n}$. $\epsilon$ is a parameter of the measurement 
inaccuracy. As $\epsilon$ tends to zero $\wne(\ve{m})$ should tend towards a Dirac 
$\delta$-function peaked at $\ve{n}$. The probability that 
such an imprecisely specified measurement yields 
an outcome +1 when the system is prepared in some pure state $\psi$ is 
\begin{eqnarray*}
\mbox{\rm Prob}_\psi^{\ve{n},\epsilon}(+1) & = & 
\int_{S^2}d\Omega (\ve{m}) \wne(\ve{m}) \tr (P_\psi P_{\ve{m} , +1}) \\
& = & \tr \left( P_\psi \int_{S^2}d\Omega (\ve{m}) \wne(\ve{m})  P_{\ve{m} , +1} \right),
\end{eqnarray*}
where $d\Omega$ is the Lesbesgue-measure of the sphere.   
Defining 
\begin{equation}
\label{def-Fne}
F^{\ve{n},\epsilon}(i) :=  
\int_{S^2} d\Omega(\ve{m}) \, w_{\ve{n},\epsilon}(\ve{m}) P_{\ve{m},i},
\end{equation}
the probability of getting outcome $i$ if the system is prepared in state $\psi$ is  
$$
\mbox{\rm Prob}_\psi^{\ve{n}, \epsilon}(i) 
=  \tr \left(P_\psi F^{\ve{n},\, \epsilon}(i)\right).
$$
From (\ref{def-Fne}) it is obvious that the $\Fne$ are positive self-adjoint operators 
satisfying $0 \leq \Fne(i) \leq \id$. But they are not projectors since 
$\Fne(i)\not= \Fne(i)^2$. 
The $\Fne(i)$ form a resolution of the identity,
\begin{equation}
\label{eq-resolution of identity}
\Fne(1) + \Fne(0) + \Fne(-1) = \id,
\end{equation}
which follows from (\ref{def-Fne}) and 
$P_{\ve{n},1} + P_{\ve{n},0} + P_{\ve{n},-1}= \id$.
Thus we have a positive operator valued (POV) measure which associates to each element $i$
of the value space $\{1,0,-1\}$ the positive operator $\Fne(i)$.
POV-measures are the standard
tool for describing inaccurate experiments \cite{Busch et al 1995, Holevo 1982}.

We say that the error distribution $\wne(\ve{m})$ transforms covariantly if
\begin{equation}
w_{R\ve{n}, \epsilon}(R\ve{m})= \wne(\ve{m})
\end{equation}
for all rotations $R$. When both the observed system and the apparatus are rotated by $R$ the measurement statistics should not change. 
\begin{proposition}
\label{prop1}
If the distribution $\wne$ of apparatus misalignments transforms covariantly under rotations then the unsharp spin properties $\Fne$ transform covariantly under rotations,
\begin{eqnarray}
\label{eq-covariance of Fne}
D^1(R) \, F^{\ve{n},\, \epsilon}(i) \, D^1(R)^{-1} = F^{R^{-1}\ve{n}, \epsilon}(i),
\end{eqnarray}
where $D^1$ is the spin 1-representation of the rotation group.
\end{proposition}
A proof of this and the following propositions can be found in \cite{Breuer 2002}.

Since the $\Fne(i)$ transform covariantly under rotations, they are angular momentum properties
and can be regarded as spin properties with the same justification as the sharp spin properties $P_{\ve{n},i}$.
This is in line with Weyl's idea of defining observables by their transformation properties under some kinematic 
group.
\begin{proposition}
\label{prop2}
If the distribution $\wne$ of measurement errors transform covariantly under rotations then the unsharp spin properties $\Fne(i)$ have the same eigenvectors as the sharp spin properties $P_{\ve{m} \, i}$. Since the sharp spin properties $P_{\ve{n},i}$ with
$i=1, 0, -1$ have simultaneous eigenvectors and commute, this is also the case for 
the unsharp spin properties $\Fne(i)$. 
\end{proposition}
\begin{proposition}
\label{prop3}
The eigenvalues of the unsharp spin properties $\Fne(i)$ are in the set $\{\alpha_1, \ldots, \alpha_4 \}$, 
\begin{eqnarray}
\alpha_1 & = & 2\pi \int_0^\pi d\theta 
\, w_{\ve{z}, \epsilon}(\theta) \sin (\theta)  \cos (\theta/2)^4  
\nonumber \\
\alpha_2  & = & \pi \,\int_0^\pi d\theta  
\, w_{\ve{z}, \epsilon}(\theta) \sin (\theta) \sin (\theta)^2 
\nonumber \\
\alpha_3  & = &  2\pi \, \int_0^\pi d\theta  
\, w_{\ve{z}, \epsilon}(\theta) \sin (\theta) \sin (\theta/2)^4 
\label{eq-eigenvalues of Fne} \\
\alpha_4  & = &  2\pi \, \int_0^\pi d\theta  
\, w_{\ve{z}, \epsilon}(\theta) \sin (\theta) \cos (\theta)^2 \nonumber
\end{eqnarray}
which are all between $0$ and $1$.  
The eigenvalues of each $\Fne(i)$ add up to $1$.
\end{proposition}

To give an explicit example, assume that the
spin directions actually measured are uniformly distributed over $\C{n}$, 
the set of directions
deviating from $\ve{n}$ by less than an angle $\epsilon$. Denoting by $A$ the area of
$\C{n}$ on the unit sphere, $\wne$ in this example is  $1/A$ times the characteristic function of 
$\C{n}$. The $F^{\ve{z}, \epsilon}(i)$ are diagonal matrices of the form 
\begin{subequations}
\label{eq-Fze}
\begin{equation}
F^{\ve{z}, \epsilon}(1) =
\left(
\begin{array}{ccc}
 \alpha_1 & 0 & 0 \\
0 &
\alpha_2 & 0 \\
0 & 0 & \alpha_3
\end{array}
\right) 
\end{equation}  
\begin{equation}
\label{eq-Fz0}
F^{\ve{z}, \epsilon}(0) = 
\left(
\begin{array}{ccc}
\alpha_2 & 0 & 0 \\
0 &
\alpha_4 & 0 \\
0 & 0 & \alpha_2
\end{array}
\right) 
\end{equation}
\begin{equation}
F^{\ve{z}, \epsilon}(-1) = 
\left(
\begin{array}{ccc}
\alpha_3 & 0 & 0 \\
0 &
\alpha_2 & 0 \\
0 & 0 & \alpha_1
\end{array}
\right), 
\end{equation}
\end{subequations}
where the $\alpha_i$ are given by 
\begin{eqnarray}
\alpha_1 & = & \frac{1}{24} \left(15 + 8 \cos(\epsilon) + \cos(2\epsilon) \right) \nonumber \\
\alpha_2 & = & \frac{1}{3} \left(2 + \cos(\epsilon)\right)  \sin(\epsilon/2)^2 \nonumber \\
\alpha_3 & = & \frac{1}{3} \sin(\epsilon/2)^4 
\label{eq-eigenvalues of Fne for uniform distribution} \\
\alpha_4 & = & \frac{1}{6} \left(3 + 2 \cos(\epsilon) + \cos(2\epsilon)\right) . \nonumber
\end{eqnarray}
Observe that, as the measurement inaccuracy $\epsilon$ tends to zero, two eigenvalues 
($\alpha_2$ resp. $\alpha_3$) of each $F^{\ve{z},\epsilon}(i)$ go to zero, 
one eigenvalue ($\alpha_1$ resp. $\alpha_4$) goes to one. The unsharp spin properties
$F^{\ve{z},\epsilon}(i)$ converge to the sharp spin properties $P_{\ve{z},i}$ as given for example in eq. (\ref{eq-Pz1}).

\section{\label{section-The KS-theorem for Unsharp Spin 1 Observables}
The KS-theorem for Unsharp Spin 1 Observables}
Determining the result $i\in\{1,0,-1\}$ of a {\em sharp} spin measurement in direction 
$\ve{n}$ is picking one of the sharp spin properties $P_{\ve{n},i}$ and 
assigning it the truth value 1. 
Since the sharp spin properties are projectors 
$P_{\ve{n},i}:=\vert\psi_{\ve{n},i}\rangle\langle\psi_{\ve{n},i}\vert$
they can be identified with the ray $\psi_{\ve{n},i}$. So, assigning the value 1 to one of 
the $P_{\ve{n},i}$ and the value 0 to the other two, is equivalent to assigning the colour
T (true) to one of the rays $\psi_{\ve{n},i}$ and the colour F (false) 
to the two other rays. 
The traditional KS-proofs show that for certain sets of directions
this colouring rule cannot be satisfied.

Determining the result $i\in\{1,0,-1\}$ of an {\em unsharp} spin measurement in 
direction $\ve{n}$ is picking one of the unsharp spin properties 
$\Fne(i)$ and assigning it the truth value 1. 
But the unsharp spin properties are not projectors and therefore cannot readily be
identified with rays. To arrive at a colouring rule for rays we have to proceed in 
a different way. Fix some unsharpness tolerance $0.5>\delta \geq 0$.
\begin{definition}
\label{def-colouring}
If the outcome of a spin measurement with some intended direction $\ve{n}$ is 
$i\in\{1,0,-1\}$, 
then the ray of the eigenvector of $\Fne(i)$ corresponding to an eigenvalue larger or 
equal to $1-\delta$ gets colour AT (almost true), and the rays of eigenvectors corresponding to 
eigenvalues smaller or equal to $\delta$ get colour AF (almost false). (Rays corresponding
to some eigenvalue between $\delta$ and $1-\delta$ are not assigned a colour.)
\end{definition} 
$\delta>0$ is an unsharpness tolerance below which an eigenvalue counts as ``almost zero".
An eigenvalue above $1-\delta$ counts as ``almost one". The exact level of $\delta$
is a matter of taste,
and our results do not depend on the exact level. But certainly $\delta$ should be smaller 
than 0.5, since otherwise some values could simultaneously be counted as almost zero
and almost one. 

An example: Assume that one intends to measure spin in direction $\ve{z}$ and
result 0 occurs. This means that the unsharp spin property $F^{\ve{z},\epsilon}(0)$ 
is realised, whereas $F^{\ve{z},\epsilon}(1)$ and $F^{\ve{z},\epsilon}(-1)$ are not 
realised. Accordingly we assign the colour AT to the ray $(0,1,0)$, 
which, by (\ref{eq-Fz0}), is the eigenvector of $F^{\ve{z},\epsilon}(0)$ 
with eigenvalue close to 1. 
To $(1,0,0)$ and $(0,0,1)$ we assign the colour AF because they are eigenvectors 
of $F^{\ve{z},\epsilon}(0)$ with eigenvalue close to 0. 
Had the outcome been +1, we would have assigned AT to $(1,0,0)$ and AF to $(0,1,0)$ 
and to $(0,0,1)$. 

In Proposition~\ref{prop2} we have seen that for a fixed intended measurement direction
$\ve{n}$ the $\{\Fne(i)\}_{i=1,0,-1}$ have the same eigenvectors. 
If the measurement inaccuracy is sufficiently small---or to me more precise: 
if the density $\wne(\ve{m})$ of apparatus misalignments has enough probability mass 
sufficiently close to the intended measurement direction $\ve{n}$---then 
the eigenvalues $\alpha_1$ and
$\alpha_4$ in eqs. (\ref{eq-eigenvalues of Fne}) will be larger than $1-\delta$, 
whereas $\alpha_2$ and $\alpha_3$ will be smaller than $\delta$. If this is the case,
exactly one ray in the orthogonal triad of eigenvectors of the $\Fne(i)$ will
get colour AT, and two rays will get colour AF. 

For example if we assume apparatus misalignments to be uniformly distributed over 
the set of directions deviating by less than an angle $\epsilon$, and if we 
choose $\delta=0.1$, 
then we can calculate from eqs. (\ref{eq-eigenvalues of Fne for uniform distribution}) 
that for $\epsilon$ smaller than 0.459 rad $ =26.3^{\rm o}$ the eigenvalues 
$\alpha_1,\alpha_4$ will be larger than $0.9$,  
while $\alpha_2, \alpha_3$ will be smaller than $0.1$. So for $\delta=0.1$, 
if the measurement inaccuracy $\epsilon$ is smaller than 0.459 rad, 
then in all orthogonal tripods one ray will be coloured AT and two rays will be 
coloured AF.       

One ray can be eigenvector of spin properties $\Fne(i), \Fme(i)$ 
in different directions $\ve{n}, \ve{m}$. Non-contextuality of the hidden variable-model 
implies that such a ray is assigned a unique colour.
Now the KS-theorem for unsharp spin observables follows in exactly the same way as the one
for sharp observables. 
In every orthogonal tripod one of the rays is constrained to
get the colour AT, the other two rays get AF. For the KS-sets 
\cite{Kochen and Specker 1967, Peres 1995, Zimba and Penrose 1993} 
of tripods such a colouring is impossible.  
Thus we arrive at 
\begin{theorem}
\label{theorem 1}
For any unsharpness tolerance $0.5>\delta\geq 0$, if in an unsharp spin 1 measurement \\
(1) the densities of apparatus misalignments transform  covariantly, 
$w_{R\ve{n}, \epsilon} (R\ve{m})=\wne(\ve{m})$, and \\ 
(2) the measurement inaccuracy described by the densities $\wne(\ve{m})$ of apparatus 
misalignments is so small that in eqs. 
(\ref{eq-eigenvalues of Fne}) $\alpha_1$ and
$\alpha_4$  are larger or equal to $1-\delta$, 
while $\alpha_2$ and $\alpha_3$ are smaller or equal to $\delta$, \\
then not all the unsharp spin observables in a KS-set of directions can consistently 
be assigned approximate truth-values in a non-contextual way.
\end{theorem}

Reading this result contrapositively we conclude that in Meyer's \cite{Meyer 1999} 
model the density of apparatus misalignments is not covariant. This can also be checked 
directly. According to his model, if we intend to measure spin in a
direction $\ve{n}$ with irrational coordinates,
the apparatus is in fact aligned in some direction $\ve{m}$ with 
rational coordinates, which is very close to $\ve{n}$. Now let $R$ be a rotation by 
$\pi/4$ around an axis orthogonal to $\ve{m}$. 
If $\wne(\ve{m})$ were rotation
covariant, in an experiment designed to measure spin in direction $R\ve{n}$ the apparatus 
would 
actually be aligned in direction $R\ve{m}$. 
But $R\ve{m}$ cannot have rational 
coordinates if $\ve{m}$ has. So $R\ve{m}$ would not be assigned a colour. 
Thus in Meyer's model the distribution of misalignments cannot be 
covariant under all rotations.

Kent \cite{Kent 1999} and  Clifton and Kent \cite{Clifton and Kent 2000} generalise
Meyer's argument to unsharp observables. They show that there are non-contextual hidden
variable models which recover the quantum probabilities of POV-measurements with
arbitrarily small inaccuracy. Reading Theorem~\ref{theorem 1} contrapositively we conclude
that the apparatus misalignments in their models are not covariant either. This can be 
checked directly on their model.

\end{document}